\documentclass[12pt,twoside,fleqn]{article}
\usepackage{epsfig}
\usepackage{amstex}
\usepackage{amssymb}

\begin{document}

\def\obs{{\cal O}}

\thispagestyle{empty}

\begin{center}
{\Large \bf {A lattice test of alternative interpretations \\[0.25cm] 
of ``triviality'' in $(\lambda \Phi^4)_4$ theory} }

\end{center}
\vspace{1.0cm}
\begin{center}
{\large 
A. Agodi$^{1,2}$, 
G. Andronico$^{1,2}$, 
P. Cea$^{3,4}$,
M. Consoli$^{1}$,
L. Cosmai$^{3}$,\\[0.125cm]
R. Fiore$^{5,6}$, and
P. M. Stevenson$^{7}$\\
\vspace{1.0cm}
{\small
$^1$~INFN - Sezione di Catania, Corso Italia 57, I 95129 Catania,  Italy \\[0.05cm]
$^2$~Dipartimento di Fisica,  Universit\`a di Catania, Corso Italia 57, I 95129 Catania, Italy\\[0.05cm]
$^3$~INFN - Sezione di Bari, via Amendola 173, I 70126 Bari, Italy \\[0.05cm]
$^4$~Dipartimento di Fisica,  Universit\`a di Bari,  via Amendola 173, I 70126 Bari, Italy\\[0.05cm] 
$^5$~INFN - Gruppo Collegato di Cosenza, I 87030 Arcavacata di
Rende (Cs), Italy\\[0.05cm]
$^6$~Dipartimento di Fisica,  Universit\`a della Calabria, I 87030 Arcavacata di
Rende (Cs),I taly\\[0.05cm]
$^7$~T. W. Bonner Laboratory, Physics Department, Rice University, Houston, TX 77251, USA\\[0.05cm]
} 
}

\end{center}
\vspace{0.8cm}
\begin{center}
{\large {\bf Abstract}}
\end{center}
\vspace{0.3cm}
There are two physically different interpretations of ``triviality'' in
$(\lambda\Phi^4)_4$ theories.  The conventional description predicts a 
second-order phase transition and that the Higgs mass $m_h$ must vanish 
in the continuum limit if $v$, the physical v.e.v, is held fixed.  
An alternative interpretation, based on the effective potential obtained 
in ``triviality-compatible'' approximations (in which the shifted `Higgs' 
field $h(x)\equiv \Phi(x)-\langle \Phi \rangle$ is governed by an effective 
quadratic Hamiltonian) predicts a phase transition that is very weakly 
first-order and that $m_h$ and $v$ are both finite, cutoff-independent 
quantities.  To test these two alternatives, we have numerically computed 
the effective potential on the lattice.  Three different methods were 
used to determine the critical bare mass for the chosen bare coupling value.  
All give excellent agreement with the literature value.  Two different 
methods for obtaining the effective potential were used, as a control on 
the results.  Our lattice data are fitted very well by the 
predictions of the unconventional picture, but poorly by the conventional 
picture.

\newpage
\setcounter{page}{1}

\section{Introduction}

   One of the most interesting results of modern quantum field theory concerns
the ``triviality'' \cite{aizen,froh,sokal,latt,luscher87,glimm,book} of  
$(\lambda\Phi^4)_4$ theories.  The physical meaning of this mathematical 
result remains controversial, however.  The conventional interpretation is 
based on Renormalization-Group-Improved-Perturbation-Theory (RGIPT), 
while a quite different interpretation is advocated in Refs. 
\cite{zeit,primer,response}.  The two pictures have drastically different 
implications for the Standard-Model phenomenology.  They also give different 
predictions for the effective potential, and the purpose of this paper 
is to compare those predictions with a model-independent lattice calculation 
of $V_{\rm eff}$.  That is, we perform a precise numerical {\it experiment} 
as a test of the two alternatives.  

\par  The conventional interpretation of ``triviality'' is as follows:  
Leading-order RGIPT predicts that the running coupling constant, if 
finite at some low energy scale, will blow up and become infinite at 
some larger energy, the Landau pole.  The only way to avoid this 
unphysical behaviour, it is argued, is to push the Landau scale to 
infinity by sending the low-energy renormalized coupling $\lambda_R$ 
to zero, thus recovering ``triviality.''  In RGIPT the Higgs mass 
$m_h$ is proportional to $(\lambda_R v^2)^{1/2}$ and goes to zero in 
the continuum limit, if the vacuum expectation value $v$ 
(phenomenologically determined to be $\sim\, 246$ GeV) is taken to be 
finite.  In this picture, the only way to have a viable Higgs mass is 
to keep the cutoff $\Lambda$ finite and not too large.  The scalar 
sector of the Standard Model is then a non-renormalizable theory in 
which the mysterious cutoff plays a crucial role.  

\par  One problem with this explanation is that RGIPT does not give a 
consistent picture \cite{landau}.  The Landau pole appearing in 
leading order is absent in next-to-leading order.  Instead, there is 
an ultraviolet fixed point $\overline {\lambda}$, due to the negative 
sign of the 2-loop coefficient of the perturbative $\beta$-function.  
Taken at face value, the two-loop result implies a finite bare coupling 
constant, with the renormalized coupling lying anywhere in the region 
$0$ to $\overline {\lambda}$.  Since there is no reason for $\lambda_R$ 
to vanish in this case, one has a direct conflict with ``triviality'' 
\cite{higher}.

\par  It is usually asserted that the leading-order RGIPT picture is 
supported by {\it non-perturbative} lattice simulations showing that 
$m_h$ vanishes in units of $v$ in the continuum limit.  While the evidence 
certainly implies $m_h/v_B \to 0$ (in accord with the rigorous results 
of sect. 15 of \cite{book} that $m_h$ and $v_B$ cannot scale uniformly 
in the continuum limit), the crucial issue is how the bare vacuum field 
$v_B$ measured on the lattice is related to the physical $v\sim$~246 GeV 
defined from the Fermi constant.  Up to now \cite{lang} one has used the 
long-distance behaviour of the propagator of the {\it shifted} field(s) 
to extract a renormalization constant $Z \equiv Z_h$, and defined 
$v=v_B/\sqrt{Z_h}$.  The lattice data provide overwhelming evidence that 
$Z_h \sim 1$, as one would expect for a trivially free shifted field.  
However \cite{zeit,primer,response}, is $v_B/\sqrt{Z_h}$ the correct 
definition of the physical {\it vacuum } field?  More precisely, if one 
considers the exact definition of the physical vacuum field from the 
effective potential, namely 
$\phi_R=\phi_B/\sqrt{Z_{\phi}}$ such that
 \begin{displaymath}
\left. \frac{d^2V_{\rm eff}}{d\phi^2_R} \right|_{\phi_R=\pm v}=m^2_h,
\end{displaymath}
does one find $Z_{\phi}=Z_h$ up to negligible corrections? 

\par  In fact, as discussed in refs. \cite{zeit,primer,response}, one 
finds $Z_{\phi} \neq Z_h$ in any approximation (e.g., one-loop, gaussian 
\cite{zeit}, or post-gaussian \cite{rit2}) that mimics the basic 
``triviality'' of the theory, where the shifted field 
$h(x)\equiv \Phi(x)-\phi_B$ is consistently described by an effective 
quadratic Hamiltonian.  In such approximations there is a non-trivial 
$Z_{\phi}$, although $Z_h=1$ holds identically.  The resulting effective 
potential, $V_{\rm triv}$, has a simple, universal structure given by 
the sum of a (suitably redefined) classical potential and a (suitably 
redefined) zero-point energy for a free field with a $\phi$-dependent mass. 
[``Suitably redefined'' refers to the usual mass-renormalization or 
normal-ordering procedure.]  The crucial point is that this simple structure 
-- originally obtained in the Coleman-Weinberg {\it one-loop} calculation 
\cite{cw} -- is a nonperturbative consequence of ``triviality''.  The fact 
that the one-loop approximation seems untrustworthy from a loop-expansion 
perspective \cite{cw} is not relevant.  The one-loop computation should 
rather be viewed as the prototype for a class of nonperturbative, 
``triviality-compatible'' computations of $V_{\rm eff}$, which all yield 
the same result, $V_{\rm triv}$.  The form of $V_{\rm triv}$ reflects the 
coexistence of spontaneous symmetry breaking (SSB) and ``triviality'', 
and becomes the basis for an alternative renormalization procedure that 
replaces the standard perturbative approach.

\par  To understand the basic difference with the usual perturbative approach, 
consider the following question: In the continuum limit of a ``trivial'' 
theory, where the S-matrix for $2 \to 2$ particle scattering reduces to 
the trivial identity, can one still generate a finite energy density to 
de-stabilize the perturbative vacuum?  Indeed, yes, and statistical physics 
is full of such examples; e.g. superconductivity, where an arbitrarily small 
2-body interaction produces macroscopic effects.  The physical mechanism 
is well known: a tiny 2-body interaction $g$ can produce drastic changes in 
the vacuum structure if there is a sufficiently large number of states 
at the Fermi surface. In this situation, where the energy gap in the 
single-particle spectrum is a many-body effect, ordinary perturbation theory 
in $g$ fails in predicting all the basic features of the superconducting 
ground state.

\par  Following this line of thought one deduces the simple physical picture
of refs. \cite{zeit,primer,response} where the (nearly) massless quanta of 
the symmetric phase condense in the zero 4-momentum state.  Even with an 
{\it infinitesimal} 2-body strength, the Bose condensation produces a 
{\it finite} gain in the energy density, leading to the instability of 
the perturbative vacuum.  The excitations of the new vacuum are non-trivially 
related to the original quanta, but they also have vanishingly small 
interactions.  Therefore, a proper renormalization procedure for 
$(\lambda\Phi^4)_4$ theory cannot be based on a vain attempt to generate 
a finite `$\lambda_R$', a concept which has no place in a ``trivial'' theory.  
Rather, the correct strategy has to be based on the physical requirement 
that the energy density associated with SSB is finite 
\cite{zeit,rit2,cast,bran,con,iban,new,rit,agodi}; i.e. one should require 
the effective potential to be cutoff independent.  A straightforward 
renormalization-group analysis of $V_{\rm triv}$ then shows that, while 
$m_h/v_B\to 0$, one gets $m_h/v=$~cutoff-independent.  Clearly, this has 
radically different implications for the Standard Model.  The non-trivial 
re-scaling of the vacuum field, $v=v_B/\sqrt{Z_\phi}$, where 
$Z_\phi \to \infty$, is quite distinct from the trivial $Z_h=1$ 
renormalization of the free fluctuation field.  This is the essential 
ingredient that represents in a quantum-field-theoretical context the 
intuitive notion of an infinitely dense \cite{drastic} Bose condensate at 
$p_{\mu}=0$ coexisting with trivially free excitations at $p_{\mu} \neq 0$.

\par  Another striking feature of $V_{\rm triv}$ is that it predicts a 
{\it first-order} phase transition \cite{zeit,response}.  By contrast, 
the RGIPT result, $V_{\rm pert}$, shows a second-order transition.  
With $V_{\rm triv}$ one sees that, as the bare-mass term $r_0\equiv m^2_B$ 
is made more and more negative, the SSB transition occurs at a value 
$r_0=r_s$ where $m$, the physical mass-gap of the symmetric phase, though 
infinitesimal in units of the corresponding $m_h$, is still non-zero. 
Only at an even more negative value of the bare mass, $r_0=r_c<r_s$, does 
one find the `Coleman-Weinberg regime' where $m=0$ identically.  However, 
by then the system is well inside the broken phase.  Since $r_c\neq r_s$ 
there is no continuum limit for non-zero values of the bare 
coupling $\lambda_0$, contrary to the prediction of leading-order RGIPT 
where $m$ always vanishes exactly at the phase transition.  Only if 
$\lambda_0 \to 0$ can one obtain a continuum limit from $V_{\rm triv}$.  
One finds that $|(r_c-r_s)/r_c| \sim \exp(-8\pi^2/3\lambda_0)$ \cite{zeit}, 
so that $r_c \to r_s$ as $\lambda_0 \to 0$, yielding a transition 
that asymptotically becomes second order, in agreement with the rigorous 
result known for the gaussian model.  The difference between $r_c$ and $r_s$, 
although exponentially small for weak bare coupling, remains crucial 
\cite{response}. Indeed, in the limit, {\it infinitesimal} variations of $r_0$ 
near the phase-transition value induce {\it finite} variations in the 
particle mass of the broken vacuum; in the energy-density difference between 
the two phases; and in the barrier between the two phases.  The problem with 
the conventional approach is that it looks at the phase transition on too 
coarse a scale -- making {\it finite} variations in $r_0$. Viewed on that 
scale the transition appears indistinguishable from a second-order phase 
transition and the fine details are not seen.  

\par  The {\it qualitative} difference between $V_{\rm triv}$ and 
$V_{\rm pert}$ means that the two pictures can be distinguished by 
a sufficiently precise lattice calculation of $V_{\rm eff}$.  That is, 
one can perform a model-independent, {\it numerical experiment} to test 
the predictions of both the conventional RGIPT picture and the alternative 
picture of refs. \cite{zeit,primer,response}.  Initial results were presented 
in ref. \cite{agodi}.  The aim of this paper is to provide more refined 
results from a precise lattice calculation of the slope of the effective 
potential. 

\section{The lattice effective potential}

     We begin by defining the (one-component) $(\lambda\Phi^4)_4$ 
theory on a lattice:  
\begin{equation}
\label{action}
   S =a^4 \sum_x \left[ {{1}\over{2a^2}}\sum_{\mu}(\Phi(x+a\hat e_{\mu}) - 
\Phi(x))^2 + 
{{r_0}\over{2}}\Phi^2(x)  +
{{\lambda_0}\over{4}} \Phi^4(x) - J \Phi(x) \right]    
\end{equation}
where $x$ stands for a generic lattice site, $a$ denotes the lattice
spacing, and $\lambda_0 >0$.  For SSB the basic quantity is the expectation 
value of the bare scalar field $\Phi(x)$ (B=Bare)
\begin{equation}
\label{phij}
      \langle \Phi \rangle _J = \phi_B(J)   
\end{equation}
in the presence of an external source whose strength $J$ is $x$-independent.
Determining $\phi_B(J)$ at several $J$-values is equivalent 
\cite{call2,huang} to inverting the relation 
\begin{equation}
\label{j}
    J=J(\phi_B)={{dV_{\rm eff}}\over{d\phi_B}}   
\end{equation}
involving the effective potential $V_{\rm eff}(\phi_B)$.  In this way, 
starting from the action in Eq. (\ref{action}), the effective potential of 
the theory is {\it rigorously} defined up to an arbitrary integration 
constant (usually chosen to fix $V_{\rm eff} (0)=0$).  [This definition 
is equivalent to the Legendre transform definition and is convex 
downward \cite{convex}.]  In this framework, SSB occurs when the function 
$\phi_B(J)$ has a non-zero limit as $J \to 0$:  
\begin{equation}
\label{eq:2.5}
\lim_{J\to 0^{\pm}}~\phi_B(J)=\pm v_B \neq 0.  
\end{equation}
One expects such behaviour for a certain range of the bare parameters 
$r_0$ and $\lambda_0$ appearing in the lattice action Eq. (\ref{action}). 
It corresponds to the effective potential having non-trivial minima 
with $V_{\rm eff} (\pm v_B) \leq V_{\rm eff} (0)$.

\subsection{Monte Carlo simulation}

For the Monte Carlo simulation of the lattice field theory described by 
Eq.~(\ref{action}) we used the standard Metropolis algorithm.  In order to 
avoid the trapping into metastable states due to the underlying Ising 
dynamics we followed the upgrade of the scalar field $\Phi(x)$ with the 
upgrade of the sign of $\Phi(x)$.  This is done according to the effective 
Ising action~\cite{Brower89} 
\begin{equation}
\label{Ising}
S_{\mathrm{Ising}}  =   J \, \sum_x \left| \Phi(x) \right| s(x) 
 \mbox{ } -  \sum_x \sum_{\hat{\mu}}
\left| \Phi(x+\hat{\mu}) \Phi(x) \right|
s(x+\hat{\mu}) s(x) \;,
\end{equation}
where $s(x) = \mathrm{sign}(\Phi(x))$. We measured the vacuum
expectation  value of the scalar field 
\begin{equation}
\langle\Phi\rangle_J={{1}\over{N_c}}\sum^{N_c}_{i=1}{{1}\over{L^4}}
\sum_x\Phi^i(x) \,,
\end{equation}
where $N_c$ is the number of the lattice configurations generated with the
action, Eq. (\ref{action}).

\par  Statistical errors are evaluated taking into account the autocorrelation 
time in the statistical sample generated by the Monte Carlo simulation. If we 
consider a generic observable ${\cal{O}}$ (function of the lattice 
configuration) the integrated autocorelation time is defined as
\begin{equation}
\label{autotime}
\tau_{\text{int}}(\obs) = \frac{1}{2}  \sum_{{t} \,=\, - \infty}^{\infty} 
\rho_{\obs \obs} (t) =
\frac{1}{2} \ +\  \sum_{{t} \,=\, 1}^{\infty} \  \rho_{\obs \obs} (t)  \,,
\end{equation}
where  $\rho_{\obs \obs} (t)$ is the normalized autocorrelation function:
\begin{equation}
\label{normautocorr}
\rho_{\obs \obs}(t)  \;=\;  C_{\obs \obs}(t) / C_{\obs \obs}(0) \,,
\end{equation}
and 
\begin{equation}
\label{autocorr}
C_{\obs \obs}(t)  \;=\;   \langle  \obs_s \obs_{s+t} \rangle   -  
\langle \obs \rangle ^2  
\end{equation}
is the unnormalized autocorrelation function.

\par  The integrated autocorrelation time depends on the parameters in the 
lattice action Eq.~(\ref{action}) and determines the statistical error in 
Monte Carlo measurements of the expectation value $\langle \obs \rangle$ 
of the observable $\obs$.  If $\bar{\obs}$ is the sample mean of the 
observable $\obs$
\begin{equation}
\label{samplemean}
\bar{\obs} \ \ = \ \ {1 \over n }\  \sum_{t=1}^n \ \obs_t   \,,
\end{equation}
the sample variance is 
\begin{eqnarray}
\label{samplevariance}
{\text{var}}( \bar \obs )  &= &
\langle \bar{\obs}^2 \rangle -  \langle \bar{\obs} \rangle^2 =
  {1 \over n^2} \ \sum_{r,s=1}^n \ C_{\obs \obs} (r-s)  
 = {1 \over n }\ \sum_{{t} \,=\, -(n-1)}^{n-1}
  \left( 1 -  {{|t| \over n }} \right) C_{\obs \obs}       \nonumber  \\[1mm]
 &\approx&  {1 \over n }\ (2 \tau_{int}(\obs) ) \ C_{\obs \obs} (0)
   \qquad {\rm for}\ n\gg \tau   \,.
\end{eqnarray}
Therefore the statistical error is given by
\begin{equation}
\label{staterror}
s= \sqrt{ \frac{K}{n} \,C_{\obs \obs} (0) }.
\end{equation}
For completely uncorrelated data $K = 2 \tau_{int}(\obs) =1$.

\par  Therefore, to obtain the statistical error we must estimate the 
factor $K$ in Eq.~(\ref{staterror}).  This can be achieved through a 
direct evaluation of the integrated autocorrelation time~\cite{Madras88}, 
or by using the ``blocking''~\cite{blocking} or the 
``grouped jackknife''~\cite{jackknife} algorithms.  We have checked that 
applying these three different methods we get consistent estimates of the 
statistical errors.

\subsection{Determination of the critical bare mass parameter}

    We chose to run our lattice simulation with $\lambda_0=0.5$.  
We then have $s \equiv 3\lambda_0/16\pi^2 \ll 1$, as needed for 
the continuum limit of refs. \cite{zeit,primer,response}.  This puts 
us in a region where {\it both} bare and renormalized couplings are small:
an excellent place to test the validity of perturbation theory.  

\par  The bare mass-squared $r_0$ needs to be close to the critical value 
$r_c$ so that the correlation length will be very large.  To determine 
accurately the $r_c$ for $\lambda_0=0.5$ we used several methods.  Firstly, 
there is an analysis by Brahm \cite{brahm} that yields
\begin{equation}
\label{breq}
r_c=-0.2240 -{{1.00\pm 0.05}\over{L^2}} \pm 0.0010,
\end{equation}
for an $L^4$ size lattice.  For L=16 this gives 
\begin{equation}
\label{res0}
r_c=-0.2279 (10) \,.
\end{equation}
Brahm made use of the L\"uscher-Weisz high-temperature-expansion results 
\cite{luscher87} and made lattice calculations of the susceptibility in 
the broken phase on lattices ranging from $4^4$ to $8^4$.  The 
susceptibility $\chi$ is defined as:
\begin{equation}
\label{suscep}
\chi=L^4 \left[ \left\langle \Phi^2 \right\rangle - 
\left\langle \Phi \right\rangle^2 \right] \,,
\end{equation}
where
\begin{equation}
\label{phi}
\Phi = \frac{1}{L^4} \sum_x  \Phi(x)  \, .
\end{equation}
One expects \cite{luscher87} that near the critical region 
$\chi^{-1} \sim (r_c-r_0)$, modulo logarithmic corrections to the 
free-field scaling law.  One can thus determine $r_c$ by extrapolation 
to vanishing $\chi^{-1}$.  Strictly speaking, this method is valid only 
for a second-order phase transition where $r_c=r_s$ and both $m$ and $m_h$ 
vanish at the phase transition.  In the case of a very weak first-order 
phase transition where 
$|r_c-r_s|/r_c \sim \exp(-1/(2s))$ the induced 
uncertainty should be negligible.

\par   To check Brahm's result we performed three different numerical 
calculations of $r_c$ at $\lambda_0=0.5$ on a $16^4$ lattice.  
We determined $r_c$ from the susceptibility Eq.~(\ref{suscep}) in both 
the broken and symmetric phases.  Our data are well described by the 
simple linear fit
\begin{equation}
\label{suscepfit}
\chi^{-1}=a |r-r_c| \,.
\end{equation}
We find in the broken phase:
\begin{eqnarray}
\label{brokenfit}
a & = & 1.964 (40)  \nonumber \\
r_c & = & -0.2270 (14) \nonumber \\
\chi^2/{\text{d.o.f.}} &=& 0.79  \,,
\end{eqnarray}
and in the symmetric phase
\begin{eqnarray}
\label{symmfit}
a & = & 2.529 (70)  \nonumber \\
r_c & = & -0.2296 (16) \nonumber \\
\chi^2/{\text{d.o.f.}} &=& 0.33  \,.
\end{eqnarray}
Our data do not show evidence of the logarithmic corrections.  As
a matter of fact we also tried the fits:
\begin{equation}
\label{logsuscepfit}
\chi^{-1}=a |r-r_c|  |\ln |r-r_c| |^\gamma \,,
\end{equation}
and found $\gamma$ consistent with zero in both cases [$0.020 \pm 0.121$ 
(broken phase), \, $ -0.003 \pm 0.133$ (symmetric phase)].
%
%
Our results for $\chi^{-1}$ together with the fits Eqs. (\ref{brokenfit}), 
(\ref{symmfit}) are displayed in Fig.~1.

\par  Our third calculation of $r_c$ was obtained from the generalized 
magnetization $\langle \Phi \rangle$ which should have the form 
\cite{Hasen89}:
\begin{equation}
\label{hasenfit}
\langle \Phi \rangle = \alpha (r_c - r)^{1/2} | \ln|r-r_c| |^\beta + \delta \,.
\end{equation}
Accordingly we fit our data and found
\begin{eqnarray}
\label{hasenresults}
\alpha &=&  1.459 (17)  \nonumber \\ 
r_c & = & -0.2278 (19)   \nonumber \\
\beta &=&  0.0064 (187) \nonumber \\
\delta & = & 0.0095 (169) \nonumber \\
\chi^2/{\text{d.o.f.}} &=& 0.27  \,.
\end{eqnarray}
The data and the fit are shown in Fig.~2.  Combining the various 
estimates in Eqs.(\ref{brokenfit}, \ref{symmfit}, \ref{hasenresults}) 
our final value of $r_c$ is:
\begin{equation}
\label{combined}
r_c=-0.2280 (9)  \,.
\end{equation}

\par  The agreement with Eq. (\ref{res0}) is excellent.  We thus have 
{\it three} independent confirmations that the Brahm's value $r_c=-0.2279$, 
extrapolated from smaller lattices by means of Eq. (\ref{breq}), represents 
a precise input definition of the `Coleman-Weinberg regime' on a $16^4$ 
lattice with the action Eq.~(1) at $\lambda_0=0.5$.  

\subsection{Determination of the effective potential}

     We have used two independent methods to compute the effective potential.  
Firstly, we ran simulations of the lattice action Eq.~(\ref{action}) for 
16 different values of the external source in the range 
$0.01\leq |J| \leq 0.70$.  In this way, as outlined in Eqs.~(2-4), 
we directly obtain the slope of the effective potential (from which 
$V_{\rm eff}$ can be obtained, up to an additive integration constant).  
Our results for $\langle\Phi\rangle_J=\phi_B(J)$ are shown in Table 1 
(errors are statistical only).
\begin{table}
\tabcolsep .2cm
\renewcommand{\arraystretch}{2}
\begin{center}
\begin{tabular}{|c|l||c|l|}
\hline
{$J$}       &    {$\phi_B(J)$}   &  {$J$}  &   {$\phi_B(J)$}   \\ \hline
-0.010      &   -0.288862 (695)  &  0.010  &   0.289389 (787)  \\ \hline
-0.030      &   -0.413565 (321)  &  0.030  &   0.414713 (376)  \\ \hline
-0.050      &   -0.488797 (296)  &  0.050  &   0.489132 (249)  \\ \hline
-0.075      &   -0.557737 (181)  &  0.075  &   0.557961 (182)  \\ \hline
-0.100      &   -0.612497 (169)  &  0.100  &   0.612865 (151)  \\ \hline
-0.300      &   -0.876352 (111)  &  0.300  &   0.876518 (95)   \\ \hline
-0.500      &   -1.03526~ (8)    &  0.500  &   1.03532~ (7)    \\ \hline
-0.700      &   -1.15518~ (8)    &  0.700  &   1.15528~ (7)    \\ \hline
\end{tabular}
\caption{Values of $\phi_B(J)$ obtained on a $16^4$ lattice at 
$\lambda_0=0.5$ and $r_0=r_c=-0.2279$. Errors are statistical only.}
\end{center}
\label{table:I}
\end{table}

\par  As an additional check of our results, we performed a calculation 
using an alternative approach to $V_{\rm eff}$ first proposed in 
ref. \cite{fukuda} and later extended in ref. \cite{oraifeartaigh}.
This second method is based on the approximate effective potential 
$U_{\rm eff} (\phi_B;\Omega)$ defined through
\begin{equation}
\label{eq:3.1}
\exp{-U_{\rm eff}(\phi_B;\Omega)}=\int [D\Phi]
~\delta \left( {{1}\over{\Omega}}\int \!d^4x~ \Phi(x)-\phi_B \right) 
\exp{-S[\Phi]} .
\end{equation}
In the limit in which the 4-volume $\Omega \to \infty$, $U_{\rm eff}$ 
tends to the exact $V_{\rm eff}(\phi_B)$ from Eqs. (\ref{phij}, 
\ref{j}).  The difference between $U_{\rm eff}(\phi_B;\Omega)$ and
$V_{\rm eff}(\phi_B)$ gives both a consistency check of our calculations and
an indication of the effects due to the finiteness of our lattice.
For our purposes, it is more convenient to compare $J(\phi_B)$, from Table 1, 
with the corresponding quantity in the alternative method \cite{oraifeartaigh}
\begin{equation}
\label{jeff}
J_{\rm eff}(\phi_B,\Omega) \equiv {{dU_{\rm eff}(\phi_B;\Omega)}\over{d\phi_B}}=
\lambda_0\langle\Phi^3\rangle_{\phi_B} + r_o \phi_B \,,
\end{equation}
where  
\begin{equation}
\label{eq:3.4}
\langle \Phi^3 \rangle_{\phi_B} \equiv {{1}\over{\Omega}}
\langle \int d^4x~\Phi^3(x) \rangle
\end{equation}
and all expectation values $\langle \ldots \rangle$ are computed holding 
$\langle\Phi\rangle=\phi_B=$~fixed.

\par  Let us give some more details about this second approach.  The constraint 
on the value of $\langle \Phi \rangle$ is implemented by updating a pair 
of sites at the same time so that $\langle \Phi \rangle=\phi_B$ remains 
constant.  Then we compute the action variation for this double change and 
accept or reject it by Metropolis algorithm.  The generation of the sites 
pair is such that at least one member of the pair sequentially spans the 
whole lattice. Afterwards, we perform an Ising update of the field signs 
(by using Eq.~(\ref{Ising}) \cite{Brower89} ) to avoid unwanted trapping also 
in this case.  After a run we use the jackknife algorithm~\cite{jackknife} 
to evaluate $\langle \Phi^3\rangle_{\phi_B}$ with its statistical error and 
use Eq.(\ref{jeff}) to find the value of $J_{\rm eff}$ and its associated 
statistical error.

\par The results from this computation of $J_{\rm eff}(\phi_B;\Omega)$ 
are reported in Table 2.  The input values of $\phi_B$ were chosen to be 
the output values from Table 1.  
\begin{table}
\tabcolsep .2cm
\renewcommand{\arraystretch}{2}
\begin{center}
\begin{tabular}{|c|l||c|l|}
\hline
{$\phi_B$} & {$J_{\text{eff}}(\phi_B;\Omega)$} & {$\phi_B$} & 
{$J_{\text{eff}}(\phi_B;\Omega)$} \\ \hline
-0.288862 & -0.00976 (7)   & 0.289389 & 0.00986 (9)   \\  \hline
-0.413565 & -0.02983 (7)   & 0.414713 & 0.03008 (9)   \\  \hline     
-0.488797 & -0.04992 (10)  & 0.489132 & 0.04999 (8)   \\  \hline
-0.557737 & -0.07485 (8)   & 0.557961 & 0.07493 (8)   \\  \hline    
-0.612497 & -0.09980 (9)   & 0.612865 & 0.09995 (8)   \\  \hline 
-0.876352 & -0.29978 (12)  & 0.876518 & 0.29992 (10)  \\  \hline
-1.035260 & -0.49991 (12)  & 1.035320 & 0.49993 (11)  \\  \hline    
-1.155180 & -0.69970 (15)  & 1.155280 & 0.69992 (14)  \\  \hline
\end{tabular}
\caption{Values of $J_{\text{eff}}(\phi_B;\Omega)$ from 
Eq.~(\ref{jeff}), as obtained with our
$16^4$ lattice at $\lambda_0=0.5$ and $r_0=r_c=-0.2279$.}
\end{center}
\label{table:II}
\end{table}

\par  To get the {\it total} statistical errors reported in Table 2 we have 
combined in quadrature the purely statistical error on $J_{\rm eff}$ at any 
fixed value of $\phi_B$ with that obtained by propagating the errors on 
$\phi_B$ reported in Table 1. To this end we have used a fitting function 
which provides an excellent fit to the data $J=J(\phi_B)$ from Table 1.  
This estimate of the total statistical error has been checked for 
consistency by performing, for a few values of $\phi_B$, two runs at 
$\phi_B \pm \delta\phi_B$, $\delta\phi_B$ being the statistical error
affecting $\phi_B$ as determined from Table 1. The two estimates give 
essentially equivalent results.

\section{Comparing theory with the lattice data}

We can now compare the lattice {\it data} in Tables 1 and 2 with the 
existing theoretical expectations.  In the case of refs.~\cite{zeit,response}, 
the predicted form (in the Coleman-Weinberg case, $r_0=r_c$, where no 
quadratic term is present in the effective potential) is:
\begin{equation}
\label{triviality}
J_{\rm triv}(\phi_B)={{dV_{\rm triv}}\over{d\phi_B}}=
\alpha \phi^3_B \ln (\phi^2_B) + \gamma \phi^3_B, 
\end{equation}
where $\alpha$ and $\gamma$ are free parameters.  (Their values are
approximation-dependent within the class of ``triviality-compatible'' 
approximations.)  

\par  The RGIPT prediction exists in various slightly different forms in the 
literature.  We have first used the full two-loop calculation of Ford and 
Jones \cite{jones} in the dimensional regularization scheme.  Their 
expression for the  effective potential, in a one-component theory, is 
($\lambda\equiv 6\lambda_0$) 
\begin{equation}
V^{\mathrm{2-loop}}(\phi_B)=V_0(\phi_B)+V_1(\phi_B)+V_2(\phi_B) \,,
\end{equation}
with
\begin{eqnarray*}
V_0(\phi_B) & = & \frac{\lambda}{4!}\phi^4_B +{{M^2}\over{2}}\phi^2_B \\
V_1(\phi_B) & = &
\frac{1}{64\pi^2}m^4_2[\overline{\ln}\frac{m^2_2}{\mu^2}-\frac{3}{2}]
\end{eqnarray*}
and
\begin{eqnarray}
V_2(\phi_B) & = & \frac{1}{256\pi^4} \frac{\lambda^2\phi^2_Bm^2_2}{8}       
\left[ 
5 + 8\Omega(1)-4\overline{\ln}\frac{m^2_2}{\mu^2}  
+ \overline{\ln}^2\frac{m^2_2}{\mu^2} \right]
+  \frac{1}{256\pi^4} \frac{\lambda m^4_2}{8}
\left[ 1-\overline{\ln}\frac{m^2_2}{\mu^2} \right]^2  \,.
\end{eqnarray}
In the above equations we have introduced
$m^2_2\equiv\lambda\phi^2_B/2+M^2$, 
$\Omega(1)\equiv{{3}\over{4}}S-{{1}\over{3}}\zeta(2)$ with
$S=1/2^2+1/5^2+1/8^2+\ldots$ while $\overline {\ln}$ includes in the
definition of the logarithm additional terms of the MS scheme. The 
prediction for $J$ follows by differentiation:  
\begin{equation}
\label{twoloop}
J^{\rm 2-loop}(\phi_B)={{dV^{\rm 2-loop}}\over{d\phi_B}} \,.
\end{equation}
In this case, the two free parameters are the scale $\mu$, and the mass 
parameter $M^2$ of the classical potential.

\par  A different version of the RGIPT prediction, which re-sums various 
terms, is given in Eq.~(242), Sect.~5.4.2, of the textbook by Itzykson 
and Drouffe (ID) \cite{drouffe}, namely 
\begin{equation}
\label{itzdrouffe}
J^{\rm ID}(\phi_B)=\frac{A\phi_B}{ 
\left| \ln \frac{|\mu|}{|\phi_B|} \right|^{1/3} } +
{{(4\pi)^2 \phi^3_B}\over{18  \ln{{|\mu|}\over{|\phi_B|}}  }} \,,
\end{equation}
where again we have two free parameters $A$ and $\mu$.  

\par  Note that in the two preceding equations we have ignored the 
distinction between $\phi_B$ and $\phi_R$.  This is justified because 
in these conventional approaches there is only one $Z$ (i.e., 
$Z_{\phi} \equiv Z_h$) and it is known from many lattice calculations that 
$Z_h$ is very close to unity (see Table II of ref.\cite{lang}).   
 
\par  The results of fitting the lattice data to the three predictions,
Eqs. (\ref{triviality}, \ref{twoloop}, \ref{itzdrouffe}) are reported 
in Table 3.  The first yields a good fit ($\chi^2/{\rm d.o.f.} \le 1$), 
while the latter two yield poor fits ($\chi^2/{\rm d.o.f.} \sim 6-10$).  
Thus, the data significantly favour the unconventional interpretation of 
``triviality'' proposed in refs. \cite{zeit,primer,response} over the 
conventional interpretation.  

\begin{table}
\tabcolsep .2cm
\renewcommand{\arraystretch}{2}
\begin{center}
\begin{tabular}{|c||l|c|l|}
\hline
 data    & $J^{\text{triv}}$~ Eq. (\ref{triviality}) & $ J^{\text{2-loop}}$~ 
Eq. (\ref{twoloop})
     & $J^{\text{ID}}$~ Eq. (\ref{itzdrouffe}) \\ \hline \hline
Table 1  & $\alpha=0.0152 (2)$       & $\mu=8.0304 (449)$         
& $|\mu|=2.70 (1) \times 10^8$         \\
         & $\gamma=0.4496 (1)$       & $M^2=-0.0025 (1)    $         
& $A=-0.0055 (3)$  \\
         & $\chi^2=\frac{15}{16-2}$  & $\chi^2=\frac{142}{16-2}$  
& $\chi^2=\frac{116}{16-2}$ \\ \hline \hline
Table 2 & $\alpha=0.0156 (2)$       & $\mu=7.9883 (455)$         
& $|\mu|=2.69 (1) \times 10^8$         \\
         & $\gamma=0.4494 (1)$       & $M^2=-0.0028 (1)$             
& $A=-0.0063 (3)$  \\
         & $\chi^2=\frac{10}{16-2}$ & $\chi^2=\frac{109}{16-2}$  
& $\chi^2=\frac{85}{16-2}$  \\ \hline
\end{tabular}
\caption{We report the values of the parameters together with the $\chi^2$ 
obtained by fitting Eqs.~(\ref{triviality}), (\ref{twoloop}), and 
(\ref{itzdrouffe}) to the data reported in Table 1 and 2.}
\end{center}
\label{table:III}
\end{table}
\section{Conclusions}

In this paper we have presented a numerical experiment to test the two 
alternative pictures of `triviality' presented in the Introduction.
To this end we have first determined the value of the critical bare mass
parameter $r_o=r_c$ that defines the `Coleman-Weinberg-regime' of 
$(\lambda\Phi^4)_4$ theory on our $16^4$ lattice for $\lambda_o=0.5$.  Using 
three different methods we have confirmed the pre-existing estimate 
$r_c=-0.2279$ obtained by Brahm \cite{brahm}.  Then, we have computed 
the effective potential, using two different methods as a control on our 
results.  The quality of the fits to the lattice data is important 
evidence for the unconventional interpretation proposed in 
\cite{zeit,primer,response}.

\vspace{0.5cm}
\begin{center}
{\bf ACKNOWLEDGEMENTS}
\end{center}
This work was supported in part by the U.S. Department of Energy 
under grant number DE-FG05-92ER40717.
\vfill
\eject

\newpage

\section*{FIGURE CAPTIONS}

\renewcommand{\labelenumi}{Figure \arabic{enumi}.}
\begin{enumerate}
\item  
The inverse of the susceptibility Eq.(\protect{\ref{suscep}}) vs. the bare mass-squared 
$r_0$, fitted by
Eq.(\protect{\ref{suscepfit}}) in the symmetric (left side) and 
broken-symmetry phase (right side).
\item 
The generalized magnetization $\Phi$  vs. the bare mass-squared $r_0$, fitted by
Eq.(\protect{\ref{hasenfit}}) in the broken-symmetry phase.
\end{enumerate}


\newpage
\begin{figure}[t]
\begin{center}
\epsfig{file=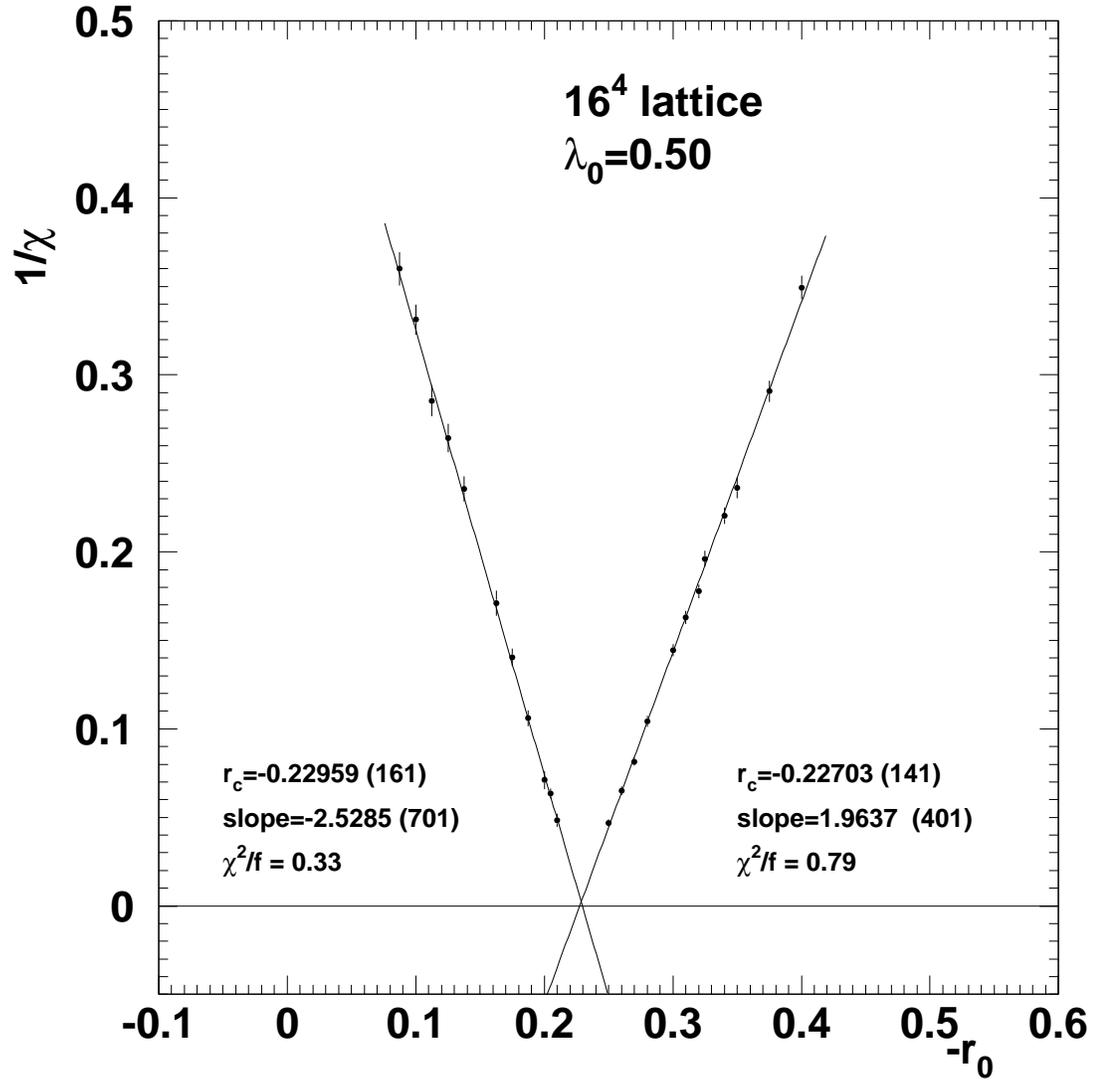,width=\textwidth}
\end{center}
\vspace{-10pt}
\caption{
The inverse of the susceptibility Eq.(\protect{\ref{suscep}}) vs. the bare mass-squared 
$r_0$, fitted by
Eq.(\protect{\ref{suscepfit}}) in the symmetric (left side) and 
broken-symmetry phase (right side).
}
\label{Fig:1}
\end{figure}

\newpage
\begin{figure}[t]
\begin{center}
\epsfig{file=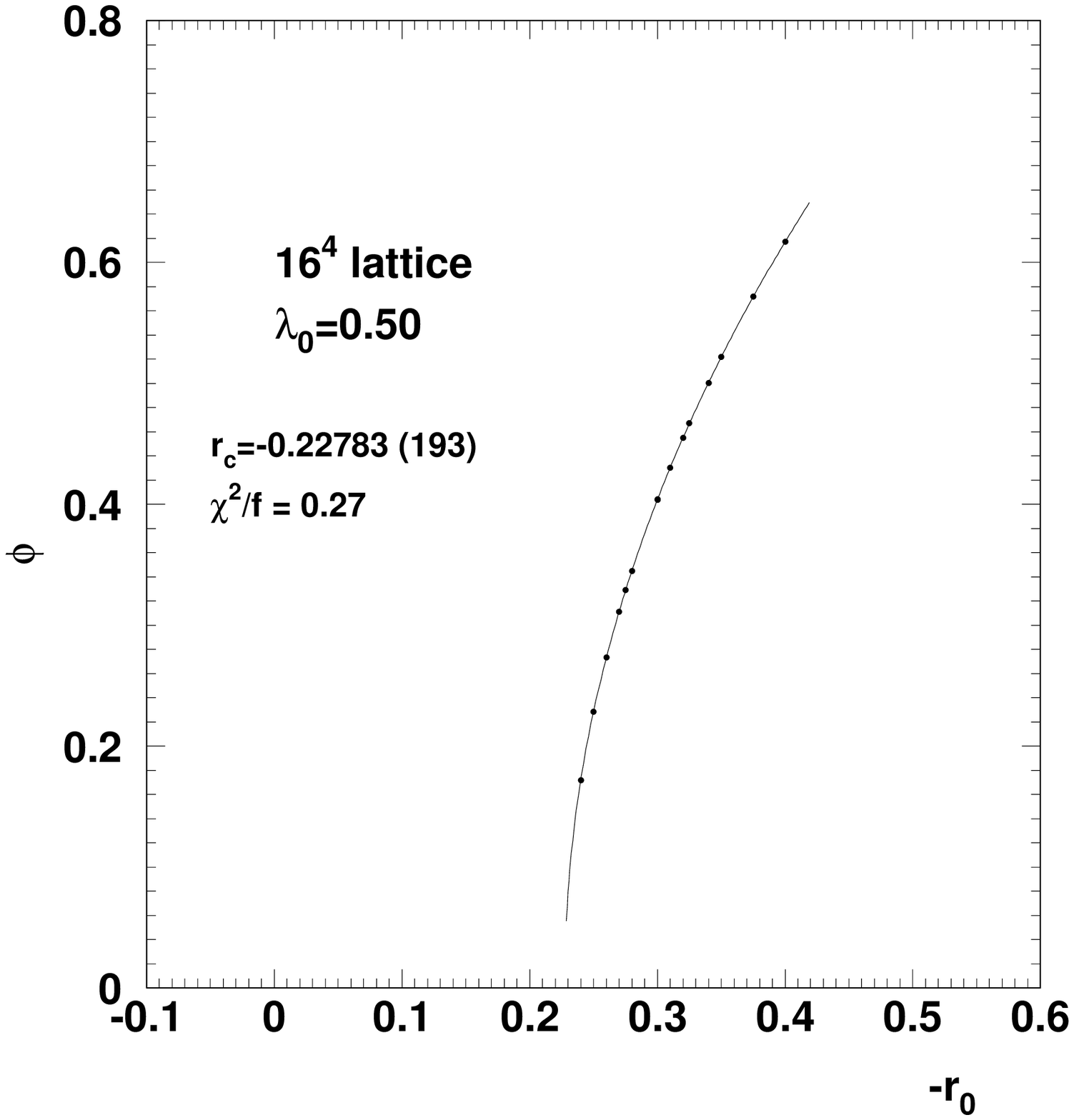,width=\textwidth}
\end{center}
\vspace{-10pt}
\caption{
The generalized magnetization $\Phi$  vs. the bare mass-squared $r_0$, fitted by
Eq.(\protect{\ref{hasenfit}}) in the broken-symmetry phase.
}
\label{Fig:2}
\end{figure}


\begin{thebibliography}{99}
\bibitem{higher}
The problem is not solved by higher orders. Perturbative calculations up to
five loops yield $\beta$-function coefficients with alternating signs, 
maintaining the conflict between even-order and odd-order results.
For more details, see ref.\cite{landau}.
\bibitem{drastic}
The phenomenon is so drastic just because the theory is ``trivial'' and there
is no finite scale where the condensation is stopped by physical short-range 
repulsion.
\end{thebibliography}
\end{document}